\documentclass[twocolumn,,
 reprint,
 superscriptaddress,
 amsmath,amssymb
]{revtex4-1}

\usepackage[dvipdfmx]{graphicx}%
\usepackage{dcolumn}%
\usepackage{bm}%
\usepackage{color}%
\usepackage{siunitx}
\usepackage{here}

\newcommand{\mysection}[1]{\vspace{-20pt}\section{#1}}
\newcommand{\mysubsection}[1]{\vspace{-8pt}\subsection{#1}}

\begin{document}

\preprint{APS/123-QED}

\title{Large-scalable fabrication of improved Bi--Te-based flexible thermoelectric modules using a semiconductor manufacturing process}
\author{Kenjiro Okawa}
\affiliation{National Metrology Institute of Japan, National Institute of Advanced Industrial Science and Technology, 1-1-1 Umezono, Tsukuba, Ibaraki 305-8563, Japan }
\author{Yasutaka Amagai}
\affiliation{National Metrology Institute of Japan, National Institute of Advanced Industrial Science and Technology, 1-1-1 Umezono, Tsukuba, Ibaraki 305-8563, Japan }
\author{Hiroyuki Fujiki}
\affiliation{National Metrology Institute of Japan, National Institute of Advanced Industrial Science and Technology, 1-1-1 Umezono, Tsukuba, Ibaraki 305-8563, Japan }
\author{Nobu-Hisa Kaneko}
\affiliation{National Metrology Institute of Japan, National Institute of Advanced Industrial Science and Technology, 1-1-1 Umezono, Tsukuba, Ibaraki 305-8563, Japan }
\author{Nobuo Tsuchimine}
\affiliation{Toshima Manufacturing Co., Ltd., 1414 Shimonomoto, Higashimatsuyama, Saitama 355-0036, Japan}
\author{Hiroshi Kaneko}
\affiliation{Toshima Manufacturing Co., Ltd., 1414 Shimonomoto, Higashimatsuyama, Saitama 355-0036, Japan}
\author{Yuzo Tasaki}
\affiliation{Toshima Manufacturing Co., Ltd., 1414 Shimonomoto, Higashimatsuyama, Saitama 355-0036, Japan}
\author{Keiichi Ohata}
\affiliation{E-ThermoGentek Co., Ltd., 102 Kujo CID bld., 13 Shimotonodacho, Higashikujo, Minami-ku, Kyoto 601-8047, Japan}
\author{Michio Okajima}
\affiliation{E-ThermoGentek Co., Ltd., 102 Kujo CID bld., 13 Shimotonodacho, Higashikujo, Minami-ku, Kyoto 601-8047, Japan}
\author{Shutaro Nambu}
\affiliation{E-ThermoGentek Co., Ltd., 102 Kujo CID bld., 13 Shimotonodacho, Higashikujo, Minami-ku, Kyoto 601-8047, Japan}

\date{\today}

\begin{abstract}
Among the several flexible thermoelectric modules in existence, sintered Bi--Te-based modules represent a viable option because of their high output power density and flexibility, which enables the use of arbitrary heat sources. We have fabricated Bi--Te-based modules with a large-scalable fabrication process and improved their output performance. The reduction in the interconnection resistance, using thick electrodes of the flexible printed circuit, significantly improves the module's output power to 87 mW/cm$^{2}$ at $\Delta T$ = 70 K, which is 1.3-fold higher than a previous prototype module. Furthermore, the establishment of the fabrication for the top electrodes by using the surface mount technology makes it possible to realize a high-throughput manufacturing of the module. Our durability tests reveal that there is no significant change in the internal resistance of the module during 10000 cycles of mechanical bending test and 1000 cycles of thermal stress test.
\end{abstract}

\maketitle
\mysection{Introduction}\vspace{-8pt}
The search for alternative power resources and recent progress in materials research\cite{He1,Snyder2,Terasaki3} has tremendously helped in improving performance of thermoelectric (TE) materials. In power generation mode, the TE materials can directly convert waste heat into electrical energy\cite{Rowe4} that is extracted from various resources such as human bodies, plants, and automobiles. In particular, the TE conversion using a flexible TE module holds promise for supplying power to the self-powered wireless sensors in the Internet-of-Things era.\cite{Siddique5}

Presently, the research on flexible TE modules is mainly focused on the fabrication of the organic and inorganic-organic hybrid-based TE modules.\cite{Du6,Suemori7,Dong8,Tian9,Satoh10} The advantage of employing organic materials is that they are flexible, low-cost, and light weight.\cite{Chen11} However, the TE conversion efficiency of these organic TE materials is still inadequate compared with that of the inorganic materials.\cite{Du6,Bahk12,Kim13,Sun14} For this reason, the flexible TE modules have been fabricated using the state-of-the-art bismuth telluride (Bi-Te) materials\cite{Goldsmid15} with various fabrication techniques including screen printing,\cite{Kim16,We17,Kim18} inkjet printing,\cite{Ou19} sputtering,\cite{Franciosoa20,Jin21,Morgan22} and multi-scanning laser lift-off process.\cite{Kim23} These techniques allow for the fabrication of flexible inorganic-material based TE modules. However, the output power density of these modules is limited to only a few mW/cm$^{2}$ at present because of the low TE conversion efficiency of the materials, and manufacturing difficulties. These flexible TE modules are constructed at a laboratory scale, which constitutes a bottleneck to using these modules in commercial products.

\begin{figure*}[ht]
\centering
\includegraphics[clip,width=8cm, angle=270]{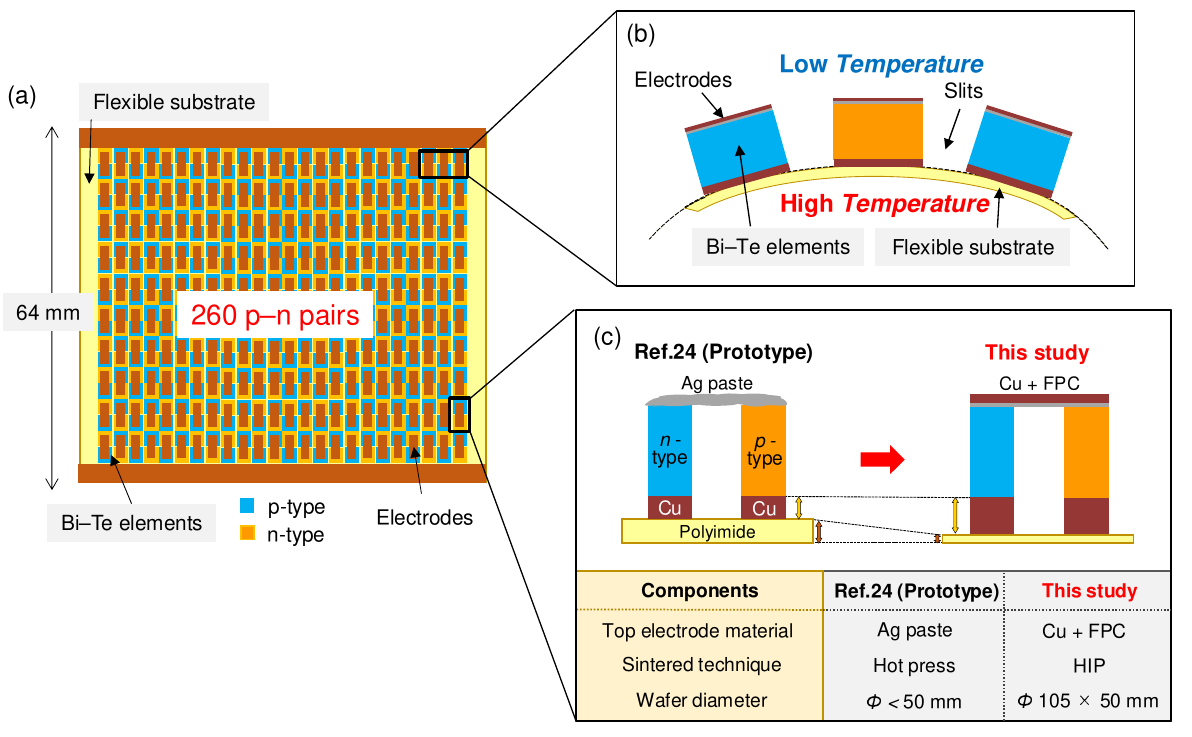}
\caption{(a) Schematic of the fabricated flexible thermoelectric module. (b) Cross-section of the module. 260~pairs of $p$- and $n$-type thermocouples of sintered Bi-Te elements are mounted on the thin flexible polyimide substrate. (c) Differences in the top electrode material, ingot sintered technique, and wafer size between this study and Ref. 24.}
\label{fig1}
\end{figure*}

Recently, a semiconductor packaging technique has been introduced for the fabrication of a flexible TE module with a unique isotropic design to address these technical issues. This report proposed a novel flexible TE module and fabrication with a semiconductor packaging technology.\cite{Sugahara24} The high-performance sintered Bi-Te chips mounted on a flexible substrate using a surface mount technology (SMT) can generate a maximum output power density of 158~mW/cm$^{2}$ at the temperature difference of $\Delta T$ = 105~K. However, owing to the high interconnection resistance, the output power of the module does not attain the value that is expected because of the intrinsic power factor of the TE materials. Moreover, the structure of the module is a prototype that cannot be mass-produced because the $p$-type and $n$-type TE materials are merely bridged with the top electrodes using Ag paste. 

In this study, we developed Bi--Te-based flexible TE modules with a large-scalable fabrication process. The designed module was achieved with a substantial reduction in its interconnection resistance (approximately 70\%) and a significant increase in its maximum output power density (1.3 times), compared with the prototype reported in Ref. 24. The fabrication process of the upper electrode was improved to enable high-throughput manufacturing of the modules. The TE conversion efficiency was acculately evaluated using the heat flow method and guarded heater method. To investigate the reliability of the module under mechanical and thermal stress, a repeated bending test with the radius as small as 15 mm and a thermal cycling test over 1000 cycles were carried out.

\mysection{Experimental methods}
\mysubsection{Fabrication of flexible thermoelectric module}\vspace{-8pt}
The flexible TE module presented in this study possesses a unique isotropic structure exhibiting uniaxial bending flexibility (Fig. \ref{fig1}(a)). In most of the conventional rigid TE modules, the $p$-type and $n$-type Bi-Te materials are arranged perpendicular to the top and bottom electrodes as the bridged $\pi$-type thermocouples. In this study, the bridged top electrodes at any location are parallel to the thermocouples unlike the conventional TE module. As all top electrodes are aligned in the same direction, the slits are created in that specific direction between adjacent thermocouples. In contrast, the bottom electrodes are vertically connected only at the end of the module. As the bottom electrodes are made up of a thin Cu film, they can withstand a limited amount of uniaxial bending stress. This design can provide a significant flexibility in the uniaxial direction (Fig. \ref{fig1}(b)). The basic design is similar to the previously reported module.\cite{Sugahara24} 

Figure \ref{fig1}(c) shows the differences in the module design and the manufacturing process between this study and Ref. 24. We improved the module design to increase the output power and reduce the interconnection resistance. To improve the thermal efficiency of the module, a thin polyimide substrate compared with that used in the previous device was used. The interconnection resistance was reduced with an increased surface area of the bottom electrode. Moreover, the fabrication process was improved to enable high-throughput manufacturing using SMT. In Ref. 24, the $p$-type and $n$-type TE materials are merely bridged with top electrodes by Ag paste because the SMT-based fabrication technique for the top electrode had not been established. This prevented the scale-up of the module fabrication. In this study, using the combined technique of flexible printed circuits (FPC) and Cu thin films, the fabrication process for the top electrode was improved to enable high-throughput manufacturing using the SMT. To prepare $\pi$-type thermocouples for the module, non-toxic materials with high TE properties, (Bi,Sb)$_{2}$Te$_{3}$ ($p$-type) and Ru-doped Bi$_{2}$Te$_{3}$ ($n$-type), were used. The high-density flexible device module comprising of 260 thermocouples in an overall dimension of $64 \times 64 \times 1.0$~mm$^{3}$ (including the flexible substrate) was fabricated, weighing approximately 9.0~g.

\begin{figure*}[ht]
\centering
\includegraphics[width=6.5cm, angle=270]{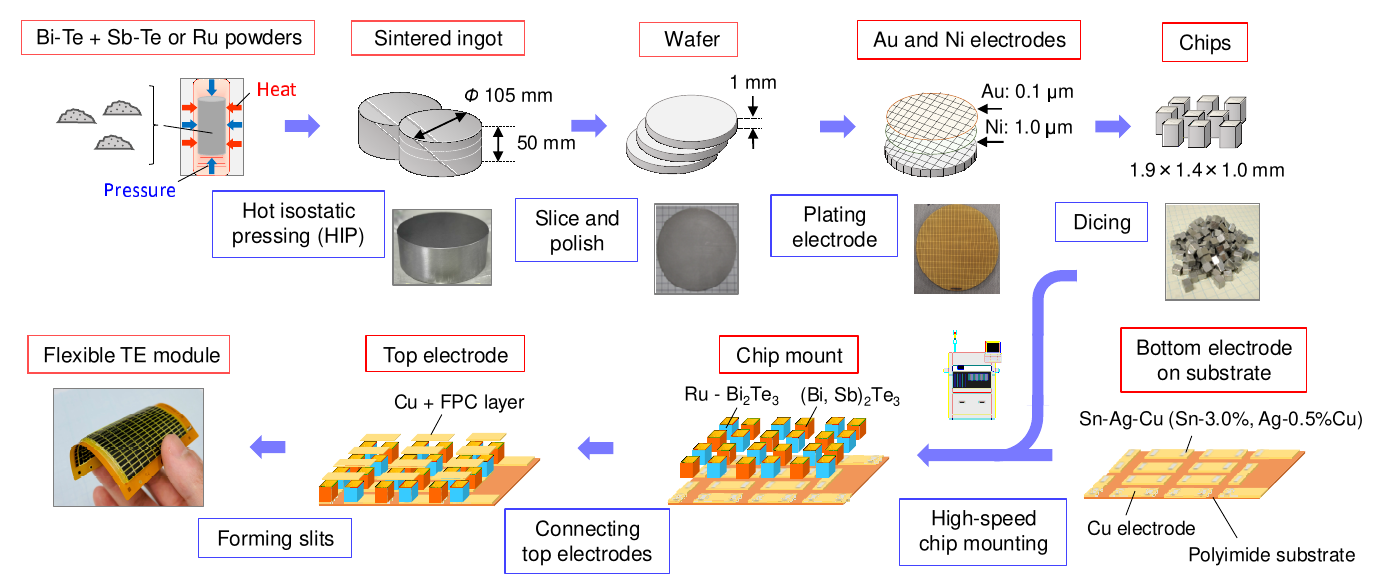}
\caption{Schematic of the fabrication process for the flexible thermoelectric module.}
\label{fig2}
\end{figure*}

Figure \ref{fig2} shows a schematic diagram of the manufacturing process used in this study. The Bi-Te and Sb-Te or Ru powders were sintered by the hot isostatic press (HIP) method to obtain ingots. The module performance may be improved with optimized grinding time and the atmosphere of the precursor powder. In Ref. 24, an ingot having $\phi$ < 50 mm was obtained using a hot-press method in which sintering was performed through uniaxial pressure. However, in this study, the HIP method, in which isotropic pressure was uniformly applied to the sample, was used. The HIP method has advantages over the hot-press method in that a higher pressure can be used, and a larger size ingot with uniform density distribution can be easily obtained because of the absence of constraints due to friction with the mold wall. A large diameter ingot ($\phi$ = 105 mm $\times$ 50 mm) was obtained using the HIP method with optimized sintering conditions. Preparation of large-scale ingots is required for the scale-up of the module fabrication. The relative density of the sintered samples—compared with the theoretical density estimated from the crystal structure—was measured to be greater than 99\%. To verify the phase purity, powder X-ray diffraction (XRD) patterns were obtained using small pieces from the densified pellets. All diffraction peaks matched with the single phase XRD peaks of Bi-Te. Our energy dispersive X-ray spectrometry and surface scanning electron microscopy results confirmed the formation of a single phase of our $p$- and $n$-type couples. The preparation of highly homogeneous large-diameter ingots contributes to improving the manufacturing yield of the modules. The electrical resistivity $\rho$ and the Seebeck coefficient $S$ of the rectangular bars cut from the ingots were measured, and the power factor (defined as $S^{2}/\rho$) was calculated (Fig. S1(a)–(c)). Both $p$- and $n$-type materials exhibit relatively high power-factors of approximately 4~mW/mK$^{2}$ at 300 K. 

The wafers ($\phi$ = 105 mm × 1 mm) were sliced from the sintered $p$- and $n$-type ingots, and thin layers of Ni (1 $\mu$m) and Au (0.1 $\mu$m) electrodes were formed by plating. For the process of electrode formation on the wafer, the 50-$\mu$m surface-polishing process introduced in this study is effective in removing cracks at the interface of the electrode. The interface between the substrate and the electrode could become a defective part when modularized. Through the polishing process, cracks entering from the surface during the wafer slicing could be removed. Moreover, no electrode peeling occurred during the dicing after the electrode formation. The small Bi-Te chips were obtained by dicing the wafer. The typical dimension of the chip was $1.9 \times 1.4 \times 1.0$ mm$^{3}$. The chips were soldered on a thin and flexible polyimide substrate. The Cu bottom electrode was mounted on the substrate. The chips and the bottom electrodes were connected using a lead-free Sn–Ag–Cu (Sn-3.0\%, Ag-0.5\%Cu)-type cream solder. The chip mounting was performed on the substrate using a high-speed chip mounting technique. The top electrode (Cu thin film combined with FPC) was fabricated on the chips using SMT. Finally, to obtain uniaxial bending flexibility, the slits were formed between adjacent thermocouples. This fabrication process enables high-throughput manufacturing of the flexible TE modules.

\mysubsection{Development of measurement apparatus}\vspace{-8pt}
The TE conversion efficiency of the modules $\eta$ is defined as $\eta = P_{\rm out}/Q_{\rm in}$, where $P_{\rm out}$ is the output power and $Q_{\rm in}$ is the input heat flux from the hot side. Generally, it is difficult to accurately estimate $Q_{\rm in}$ because there are several heat losses that are not negligible; therefore, an accurate heat flow measurement technique is required for the $\eta$ estimation.\cite{Wang25} The apparatus developed in this study utilizes heat flow method and guarded heater method to accurately measure the output power and the heat flow in a single system.

\begin{figure*}[ht]
\centering
\includegraphics[width=4cm,clip, angle=270]{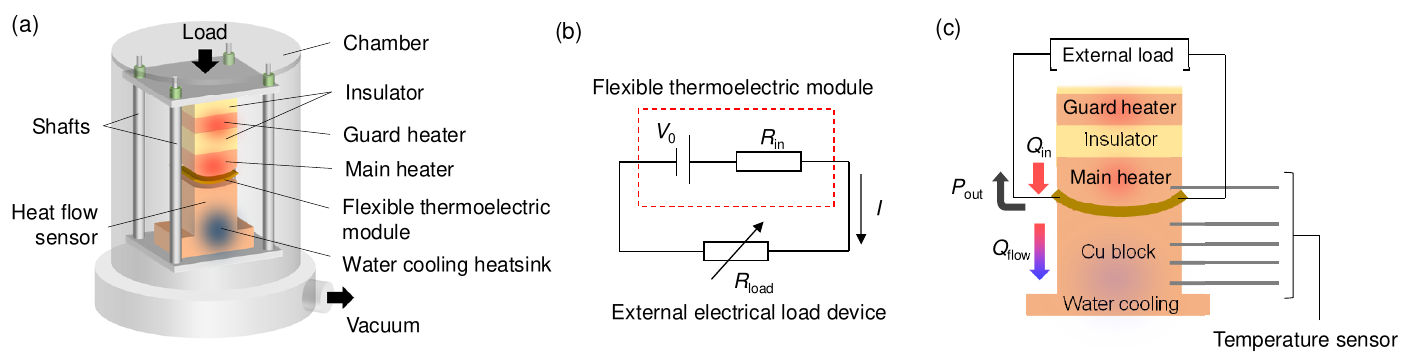}
\caption{(a) Schematic of the measurement apparatus developed in this study, and (b) simplified electrical circuit to obtain current-voltage curves of the thermoelectric module. The red dotted rectangle indicates the equivalent electrical circuit of a thermoelectric module. (c) Schematic of the measurement for the input heat flow ($Q_{\rm in}$), the heat flow from the cold side of the module ($Q_{\rm flow}$), and the output power ($P_{\rm out}$).}
\label{fig3}
\end{figure*}

 Figure \ref{fig3}(a) displays the schematic of the experimental setup developed in this study. To generate the temperature difference across the flexible TE module, a heating and cooling system was set up. The temperature of the hot side of the module could be varied in a temperature range of 303–373 K by a heater. The temperature of the cold side of the module was controlled by a water-cooling heat sink that was fixed at 303 K. Figure \ref{fig3}(b) shows the simplified circuit diagram to obtain a current-voltage curve along with the power curve. The current-voltage characteristics were measured as a function of the load current by connecting an external electrical load, and the power curve was calculated from the measured current and voltage at an applied temperature difference. The maximum output power is obtained when the load resistance $R_{\rm load}$ is equal to the internal TE module resistance $R_{\rm in}$ ($R_{\rm load} = R_{\rm in}$). The actual temperatures of the hot and cold ends were measured using thermocouples embedded in the Cu block (Figure \ref{fig3}(c)). When the heater block and the heat sink are bent into an arch, the TE conversion efficiency of the bent modules can be evaluated as well. 

The heat flow was measured by two conventional measurement techniques, the heat flow method and the guarded heater method.\cite{Wang25,Rauscher26,Takazawa27} In the heat flow method, a heat flow meter made of calibrated Cu block was used to measure the heat flow from the cold side of the module $Q_{\rm flow}$. Four thermocouples were embedded along the metal Cu block to measure the temperature gradient accurately. The heat flow can be calculated using the Fourier equation $Q$ = $\kappa A$d$T$/d$x$, where $Q$ is the heat flux, $\kappa$ is the thermal conductivity, $A$ is the cross-sectional area, and d$T$/d$x$ is the temperature gradient. $Q_{\rm flow}$ is obtained using the dimensions of the Cu metal block. The input heat flow $Q_{\rm in}$ into the module can be determined together with the maximum output power $P_{\rm out}$ generated by the module using the external electrical load device, which is calculated as $Q_{\rm in} = Q_{\rm flow} + P_{\rm out}$. Therefore, the TE conversion efficiency $\eta$ may be written as \begin{eqnarray}
\eta = \frac{P_{\rm out}}{Q_{\rm flow}+P_{\rm out}} \times 100.
\label{eq1}
\end{eqnarray}

Additionally, the guarded heater method also is used to measure $Q_{\rm in}$ using the heater power and building a stack of a polymer insulating block and a guard heater to keep the heat flow only in one direction. Assuming that all the input heat conducts through the module, $Q_{\rm in}$ can be estimated as the input power of the heater $P_{\rm heater}$ ($Q_{\rm in}$ = $P_{\rm heater}$). Thus, $\eta$ may be expressed as\begin{eqnarray}
\eta = \frac{P_{\rm out}}{P_{\rm heater}} \times 100.
\label{eq2}
\end{eqnarray}

In both methods, a special attention was paid for minimizing errors by maintaining a good thermal contact between the module samples and heater block and reducing heat loss from the heater block and heat sink with embedded heat flow meters before measuring $\eta$.\cite{Wang25,Chao28} To ensure a good thermal contact, thermal grease was applied between the copper block and the module. To further reduce the thermal resistance between the module and the copper block, the output power was measured with several different loads. A threshold value of approximately 1000 N was observed in this measurement, where the output voltage from the module did not depend on the applied load. The entire module was set in a vacuum chamber to suppress the convective heat losses.
\begin{figure*}[ht]
\centering
\includegraphics[width=11.5cm, angle=270]{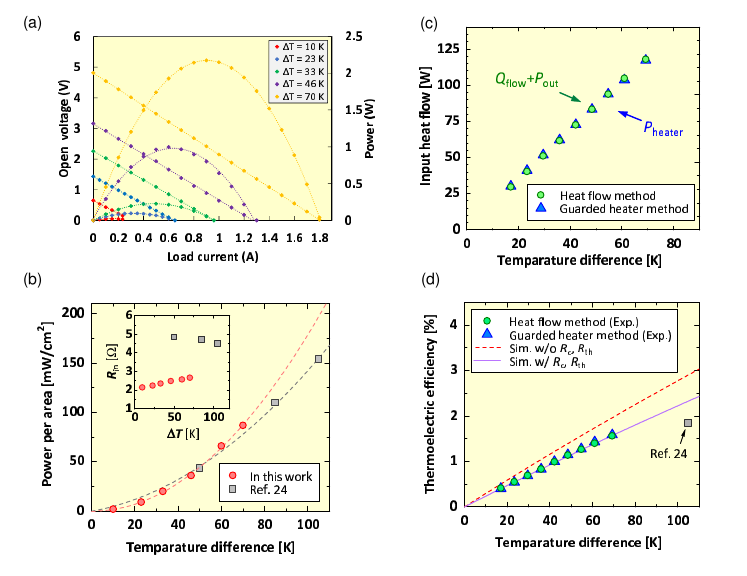}
\caption{(a) Current-voltage curves and output power for various temperature differences $\Delta T$ across the module. (b) Maximum output power per unit area and per unit weight as a function of $\Delta T$. The inset shows the internal electrical resistance $R_{\rm in}$ obtained from the current-voltage curves as a function of $\Delta T$. The red circle denotes the value obtained in this work, and the grey square represents the value reported in Ref. 24. The dotted lines are a quadratic fit that correlates the power per area and $\Delta T$.
(c) Input heat flow measured by the heat flow method (the sum of the heat flow from the cold side of the module $Q_{\rm flow}$ and the maximum output power generated by the module $P_{\rm out}$) and the guarded heater method (the input power of the heater$P_{\rm heater}$). (d) Maximum thermoelectric conversion efficiency obtained using the heat flow and guarded heater methods. The red dotted line represents the curve predicted the model based on Eq. (3), which considers only the TE materials' properties, and the solid purple line represents the curve predicted from the model, which considers the electrical contact resistance $R_{\rm c}$ and thermal contact resistance $R_{\rm th}$. The grey square indicates the value reported in Ref. 24.}
\label{fig4}
\end{figure*}
\vspace{8pt}
\mysection{Results and discussion}\vspace{-8pt}
Figure \ref{fig4}(a) shows the measured current-voltage curves and the calculated output power obtained from the measured currents and voltages for various temperature differences $\Delta T$ across the module. The measured open circuit voltage of the flexible TE module was obtained as 4.9~V and the maximum output power as 2.2~W at $\Delta T$ = 70~K. The measured open circuit voltage was consistent with the values calculated from the measured Seebeck coefficients of both $p$- and $n$-type materials (see supplementary information, Fig. S1(b)). The maximum output power per module area and per module weight are shown in Fig. \ref{fig4}(b). The maximum output power per module area increased with the temperature difference across the module and reached a maximum value of 87~mW/cm$^{2}$ at $\Delta T$ = 70~K, and the flexible TE module (total weight = 9.0 g) generated the maximum value of output power per weight as 240~mW/g at $\Delta T$ = 70~K. The output power densities normalized by the squared temperature difference $\Delta T^{2}$ were estimated to be approximately 18~$\mu$W/cm$^{2} \cdot$K$^{2}$ and 49 $\mu$W/g$\cdot$K$^{2}$, which show over a two-fold improvement compared with that of the previously reported high-performance flexible TE modules based on inorganic materials.\cite{Kim23} As can be seen from Fig. \ref{fig4}(b), the maximum output power density showed a quadratic $\Delta T$ dependence ($\propto$ $\Delta T$$^{2}$). The grey squares indicate the values obtained from Ref. 24. In both this study and Ref. 24, the values can be well fitted by quadratic fit lines. Therefore, the heat loss, such as radiation and convection, does not significantly affect the measurement of the output power, and it can be expected that accurate estimation is conducted. Compared with the maximum output power density reported in Ref. 24 (approximately 78~mW/cm$^2$ at $\Delta T$ = 70~K), our module achieved an improvement in its maximum output power density. Moreover, at the temperature difference $\Delta T$ = 105 K, our module is expected to achieve a value of 198~mW/cm$^{2}$, which is a 1.3-fold increase and the highest value among currently existing flexible TE modules. This improvement can be explained as follows. The internal electrical resistance of the module as a function of $\Delta T$ was estimated from the slope of the current-voltage curves, as shown in the inset of Fig. \ref{fig4}(b). The internal electrical resistance was calculated as 2.6~$\Omega$ at $\Delta T$ = 70 K. The internal resistance reported in Ref. 24 is 4.7~$\Omega$, indicating that a significant reduction has been achieved in this study. As the resistance predicted from the resistivity of the TE materials is approximately 1.8~$\Omega$, an approximately 70\% reduction in the interconnection resistance is considered to have been achieved, even with the difference in the mounting density. It is unlikely that the significant improvement in the power factor has been achieved because we confirmed the thermoelectric property of the Bi-Te material obtained by hot-press method used in Ref. 24 was similar to that obtained by HIP method. Therefore, the high-performance in our TE module may be attributed mainly to the low-interconnection resistance using an improved manufacturing process, such as improving the top-electrode-manufacturing process and increasing the surface area of the bottom electrode. 

To evaluate the TE conversion efficiency of the module, we measured the input heat flowing into the module. For the measurement of the conversion efficiency, the different sample with the identical design was used. Figure \ref{fig4}(c) shows the measurement results of the input heat flow by the heat flow method as well as the guarded heater method as a function of $\Delta T$. After evaluating the input heat flow and the maximum output power, the maximum TE conversion efficiency ($\eta$) is determined using Eq. \ref{eq1}. The calculated $\eta$-values of the flexible TE module increases with an increase in the temperature difference $\Delta T$, as shown in Fig. \ref{fig4}(d). The $\eta$-value was estimated to be 1.6$\%$ using the heat flow method, which is consistent with the values obtained using the guarded heater method as well as the theoretical value obtained using Eq \ref{eq2}. As evident from Fig. \ref{fig4}(d), the difference between the mean $\eta$-values obtained using the guarded heater method and the heat flow method was less than 0.1$\%$, which is within the uncertainty in the repeated measurements. To the best of our knowledge, the TE conversion efficiency of our flexible TE modules reported in this study is the highest among the currently existing flexible TE modules. 

In general, the maximum TE conversion efficiency of a TE material is
\begin{eqnarray}
\eta = \frac{T_{\rm H}-T_{\rm C}}{T_{\rm H}}\frac{\sqrt{1+Z\bar{T}}-1}{\sqrt{1+Z\bar{T}}+T_{\rm C}/T_{\rm H}} \times 100,
\label{eq3}
\end{eqnarray}
where $T_{\rm H}$ is the hot-side temperature; $T_{\rm C}$ is the cold-side temperature; $Z (= S^{2}/\rho\kappa$) is the figure of merit, and $\bar{T}$ is the mean absolute temperature.\cite{Rowe4} The red dotted line in Fig. \ref{fig4}(d) is the line predicted by Eq. \ref{eq3}, which is based on a material's TE properties. Here, the measured $n$-type material's properties ($S_{\rm n}$ = 178.7~$\mu$V/K, $\rho_{\rm n}$ = 0.80~m$\Omega$cm) and $p$-type material's properties ($S_{\rm p}$ = 157.6~$\mu$V/K and $\rho_{\rm p}$ = 0.99~m$\Omega$cm) were used as $S = (S_{\rm n}~+~S_{\rm p}$)/2 and $\rho$ = ($\rho_{\rm n}$~+~$\rho_{\rm p}$)/2. The typical value of the thermal conductivity of a Bi-Te material was used during the calculations ($\kappa$ = 2.2~W/mK). All properties were assumed to be temperature independent. The TE conversion efficiency of the module cannot actually be explained by Eq. \ref{eq3}, because the contact electrical resistance $R_{\rm c}$ and thermal resistance $R_{\rm th}$ occur at the interface between the electrodes and substrate or material chips. Min proposed the model considering the $R_{\rm c}$ and $R_{\rm th}$.\cite{Rowe29} When the $R_{\rm c}$ and $R_{\rm th}$ are used as fitting parameters, the result of fitting by the model is shown by the solid purple line in Fig. \ref{fig4}(d). The grey square in Fig. \ref{fig4}(d) shows the value reported in Ref. 24 (1.84\% at $\Delta T$ = 105~K). The obtained experimental values in Fig. \ref{fig4}(d) are close to the line predicted by Eq. \ref{eq3}. It is considered that our module approximated the intrinsic TE performance of the mounting materials, owing to the reduction in the interconnection resistance. Furthermore, by using the thin flexible substrate with low heat loss, the heat flow was efficiently transmitted to the TE-material chips.

\begin{table}[t]
\caption{Summary of the measured thermoelectric parameters of the flexible thermoelectric module at $\Delta T$ =  70 K.}
\label{t1}
\renewcommand{\arraystretch}{1.3}
\begin{tabular}{cc}
\hline
\multicolumn{1}{c}{Parameters} &  \multicolumn{1}{c}{Values} \\[2pt]
\hline
Open circuit voltage   & 4.9 V  \\
Maximum output power density per unit area  & 87 mW/cm$^{2}$ \\
Maximum output power density per unit weight  & 240  mW/g  \\
Internal electrical resistance  &  2.6 $\Omega$  \\
Thermal resistance  &  0.6 K/W \\
Thermoelectric conversion efficiency  & 1.6 $\%$  \\
\hline
\end{tabular}
\end{table}

\begin{figure*}[ht]
\centering
\includegraphics[width=11cm, angle=270]{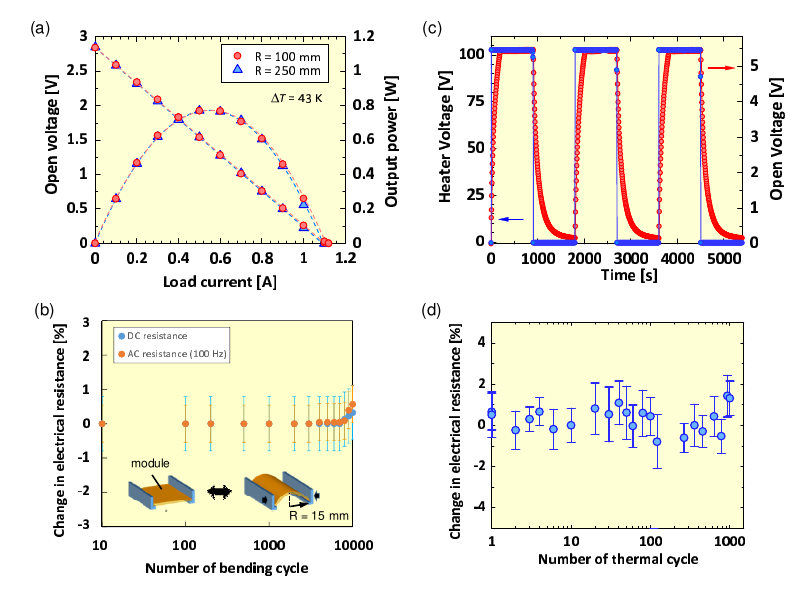}
\caption{The durability tests after subjecting the flexible TE module to the bending and thermal cycles. (a) Results of the output voltage and power for the bent states ($R$ = 100 and 250 mm). (b) Changes in the internal electrical resistance of the module are shown during  the bending cycle while bending the module along one direction with a bending radius of 15 mm. (c) Typical behavior of heater voltage and open voltage in the thermal cycling test. (d) Changes in the internal electrical resistance of the module are shown during the thermal cycling for a temperature difference of 70 K.}
\label{fig5}
\end{figure*}
The thermal resistance is defined as the ratio of the input heat flow to the resulting temperature difference in the module. The estimated thermal resistance of our flexible TE module was 0.6 K/W at $\Delta T$ = 70 K. The measured parameters of the flexible TE module at $\Delta T$ = 70 K are summarized in Table \ref{t1}.

To evaluate the durability of the module under mechanical and thermal stress, we measured the changes in the internal electrical resistance by bending our TE module along one direction as well as during a thermal cycling test. The measurement results for the bent state are shown in Fig. \ref{fig5}(a). There was no significant change in the output characteristics for both bending radius $R$~= 100 and 250 mm. The flexible TE module with an initial internal electrical resistance of approximately 3.9~$\Omega$ at $\Delta T$ = 70~K was tested in the bending test with a bending radius of 15~mm, and an ac current was applied to suppress the changes in the impedance and variations caused by the Peltier effect during the measurements. The results of the bending test are shown in Fig. \ref{fig5}(b). For a durability test, the flexible TE module was repeatedly bent up to 10000 cycles. The module shows an excellent stability along the bending direction during the test, where the measured change in the internal electrical resistance was less than 1$\%$ after 10000 cycles. The high durability against bending test originates from the asymmetrical structure, which substantially suppresses the mechanical stress along the bending direction. Additionally, the thermal cycling test was carried out with a computer-controlled dc power source that supplies the electric power to the heater and an external electric load. The typical behavior of the heater voltage and the open circuit voltage of the module during the thermal cycling test is shown in Fig. \ref{fig5}(c). The temperature of the cold side of the module was kept fixed at 303 K and that of the hot side was changed in the range of 303–373 K. The heater switch (ON/OFF) was inverted every 900 s after the temperature difference across the module was sufficiently stabilized. The change in the initial internal resistance measured during the thermal cycling test is shown in Fig. \ref{fig5}(d). The error bar denotes the standard deviation of the mean values obtained from the repeated measurements. There was no noticeable degradation up to 1000 cycles during the thermal cycling test with 2$\%$ standard deviation of the mean values calculated for the change in the internal resistance. 

\mysection{Conclusions}\vspace{-8pt}
In this study, a high-performance sintered Bi--Te-based flexible TE modules with large-scalable fabrication were developed. Improvement of the top-electrode design and increasing the surface area of the bottom electrode allows pump-up the output-power density 1.3-fold as well as TE conversion efficiency of 87 mW/cm$^{2}$ and 1.6$\%$ at $\Delta T$ = 70 K compared with a previous prototype module because of significant reduction of the interconnection resistance. The higher density mounting of thermocouples and the thinning of the flexible substrate also contributed to the improvement of the module performance. Moreover, combined with the FPC and Cu thin-film fabrication technique, the top-electrode fabrication was improved to enable high-throughput manufacturing using the SMT. The mechanical and thermal stress tests performed on these flexible TE modules showed no noticeable changes in the internal electrical resistance even under 10000 cycles of mechanical bending along one bending direction and 1000 thermal cycles at the temperature difference of 70 K. Taken together, our results provide a new opportunity for the mass production of the low-cost, high-efficiency flexible TE modules. These modules hold a promise as an alternative power source for the wireless sensors in Internet-of-Things era.

\mysection{acknowledgments}\vspace{-8pt}
This work was supported by a Leading-Edge Industry Design Project from Saitama Prefecture and a Grant-in-Aid for Research Activity Start-up (Grant No. 17H07399) from Japan Society for the Promotion of Science (JSPS).

\end{document}